\documentclass[prb,preprint]{revtex4-1} 

\usepackage{amsmath} 
\usepackage{amsfonts} 
\usepackage{graphicx} 
\usepackage{color}
\usepackage{soul}
\usepackage{multirow}

\usepackage[utf8]{inputenc} 
\usepackage[unicode=true,
bookmarks=true,bookmarksnumbered=false,bookmarksopen=false,
breaklinks=false,pdfborder={0 0 0},pdfborderstyle={},backref=false,colorlinks=true]
{hyperref}
\hypersetup{
	colorlinks=true,linkcolor=blue,citecolor=blue}

\begin{document}


\title{Motion analysis of kinetic impact projectiles for physics education in real context}

\author{Carla V. Hernández}
\email{carla.hernandez.s@usach.cl} 
\altaffiliation[permanent address: ]{Av. Ecuador 3493. 
 Santiago, Chile} 
\affiliation{Departamento de F\'isica, Universidad de Santiago, Chile}

\author{Mauricio A. Echiburu}
\email{mechiburu@uvm.cl}
\affiliation{Departamento de Ciencias Básicas, Universidad Viña del Mar, Chile}

\author{Fernando R. Humire}
\email{f.humire@academicos.uta.cl}
\affiliation{Departamento de F\'isica, Facultad de Ciencias, Universidad de Tarapac\'a, Chile}

\author{Edward F. Mosso}
\email{emosso@uta.cl}
\affiliation{ Departamento de F\'isica, Facultad de Ciencias, Universidad de Tarapac\'a, Chile}

\date{\today}

\begin{abstract}

The article presents a proposal to contextualize the study of movement in first courses of university physics, as a contribution to decision-making in situations of a social nature. For this, the case of the use of kinetic impact projectiles and the actual data provided by official sources is considered. This information is used in an object motion model describing the kinematic characteristics of a spherical projectile (a rubber bullet). For these purposes, a number Reynolds $ Re\gg 1 $ was used, which allows applying a nonlinear motion equation to find the velocity and impact energy per unit area of projectile. Results and analysis of this model can generate an interesting discussion in the classroom about the need to build protocols for the use of kinetic impact projectiles, and the importance of using scientific knowledge in social conflicts.

\end{abstract}

\maketitle 

\section{Introduction and Context} 


Usually, introductory physics courses teach little contextualized content to everyday life and disconnected from other disciplines. However, physics contributes significantly to understanding phenomena and situations related, for example, with the human body and its care\cite{Couch}. Contexts near the student can be a motivation for learning as well as developing the skills of solving own quantitative discipline problems.
Although conceptual understanding and problem-solving in physics are fundamental, current scientific education promotes the development of skills that allow addressing interdisciplinary issues, which require the search and evaluation of evidence to make sense of the information students receive from various sources\cite{Tanenbaum}. Precisely, the information that has circulated internationally in recent months has shown Chile as the scene of a civil revolution. In this context, kinetic impact projectiles, commonly known as rubber bullets, have been used to dissolve protests as in several countries\cite{INCLO}.

As a result of the use of projectiles, some people are injured in different areas of the body with varying degrees of vulnerability. Some of these injuries have resulted from the use of kinetic impact projectiles that have impacted the ocular globe causing partial or total loss of vision. Besides, the impact can produce extensive corneoscleral lacerations with either prolapse or loss of the intraocular contents requiring an early excision of the eye. In other circumstances, when the projectile has low energy, the impact can cause severe trauma injuries; but it is more likely the possibility of saving the eye\cite{Kuhn}.

According to various reports, the limit on the kinetic energy riot ammunition must have $122$ $J$ of kinetic energy\cite{Omega2000,PattenCommissionReport}. Researchers have reported that the impact energies below $20.3$ $J$ are of low risk as long as the projectile is large enough not to perforate the eye. Between $40.7$ $J$ and $122$ $J$ is considered an energy range of dangerous impact, and for impacts above $122$ $J$ would be a region of severe damage. However, other factors influence the potential damage from a shot to the human body, such as the separation range between the weapon and the subject, size, structure of the projectile, and shutter speed. Therefore, it is challenging to determine unique values for regulation of use\cite{Gobinet2011}.

In Chile, a tests performed in 2012\cite{CIPER} was conducted cartridges with 12 rubber bullets of 8 mm in diameter at distances between 5 and 30 meters. It was concluded that there is a clear possibility of serious injury generate in the human body between 5 and 25 meters away, including eye burst. Over 30 meters, slight injuries would be generated but could still involve the loss of the eye. Moreover, the experts warned that a larger dispersal distance of the pellets could affect more than one person. In the 30-meter shot, only 2 of the 12 balls hit the target. Furthermore, according to data published by the institution responsible for the study, the report stated that the speed of the projectile is $380$ $m/s$, which corresponds to supersonic speed in the conditions studied. Also, the weight of a pellet is $0.64$ $g$. In turn, the supplier indicated via a statement that the shutter speed is $320$ $m/s$. Both values are considered a reference for the present study.

Based on the background, this article proposes an analysis from physics for the kinetic impact projectile in the human eye, contextualized with real data from the Chilean case. The objective is to contribute to citizen education in introductory courses in university physics, valuing the contribution that the discipline can make to social impact discussions. With the results of the analysis, the use of basic notions of classical physics can be provided to provide evidence that favors decision-making about health care and citizenship integrity.

\section{The case of ocular damage by impact of pellets}

Several studies have investigated about eye damage from impacts with rubber bullets. In some cases, it has conducted experiments with animals to define an eye injury criterion; monkeys en Wiedenthal (1964)\cite{Wiedenthal}, pigs in Wiedenthal \& Schepens (1966)\cite{WySchepens} and Delori et al. (1969)\cite{Delori}, and also with human cadavers in Duma \& Crandall (1999)\cite{Duma_Crandall}.

In Lavy et al. \cite{Lavy2003}, the authors state that when the eye is shot a rubber bullet, it is very likely to become permanently injured by losing the eye. The authors analyzed the case of 42 patients injured in the eyes due to the use of rubber shells, of which 54\% had cutaneous lacerations, 40\% hyphema, 38\% the fractured ocular globe, 33\% an orbital fracture (bones surrounding the eye), 26\% damage to the retina, and in 21\% of the cases the projectile remained inside the eye.

Further investigations have reported the possibility of an eye injury in terms of the distance of the shot \cite{Sutter}. From a distance of 20 meters, there is a 35\% chance of hurting someone in the body and 2\% of hitting the eyes; at 10 meters away the probability is 50\%, with a 4\% of probability of hitting the eyes; and 5 meters away the probability is 80\%, with a 9\% probability of hitting the eyes.

Moreover, a study of pig eyes allowed evaluating eye hazards by measuring intraocular pressure during the impacts of high-velocity projectiles \cite{Duma2012}. Although their tests were conducted with low speeds (range between $6.2$ $m/s$, and $66.5$ $m/s$) compared to those of a riot gun, they managed to establish risk probabilities, including globe rupture depending on the diameter of the projectile. The authors investigate the correlation between intraocular pressure and normalized energy, defined as the kinetic energy divided by the cross-sectional area of the projectile. The results showed that the smaller the diameter of the projectile, the higher the probability of generating severe eye damage. It further states that a 50\% risk of globe rupture occurs just over $36000$ $J/m^2$, 50\% risk of retinal damage over $17979$ $J/m^2$, 50\% of lens damage over $17300$ $J/m^2$, and a 50\% risk of hyphema over $11700$ $J/m^2$. With these reference values,a comparison with the results obtained by analyzing the problem of impact on known real conditions can be promoted. Thus, students may offer recommendations for the use of such munitions, ensuring there is minimal risk for potential ocular damage.

\section{Projectile motion model and analysis}
The projectile motion model describes the kinematic characteristics of a spherical bullet submerged in air. As a first approach, the model uses the Reynolds number $Re$; said dimensionless number appears in many cases related to the fact flow that can be considered laminar (small Reynolds number, $Re\ll 1$) or turbulent (large Reynolds number, $1 \ll Re \lesssim 10^{5}$). These propositions allow viewing the inertial and viscous forces present in a fluid; thus, by relating the density, viscosity, velocity and typical dimension of a flow in a dimensionless expression; the Reynolds number is given by
\begin{equation}
{Re}=D V \frac{\rho}{\mu}\label{Eq:Reynolds_number}
\end{equation}
where $\rho$ is the density of the medium, $V$ is the terminal velocity of the body in the fluid, $\mu$ accounts for the viscosity of the fluid, and $D$ represents the characteristic length scale of the object in the cross-sectional plane \cite{Parker}. Moreover, since the quotient between the density and viscosity of the air, $\nu=\rho/\mu$, is of the order $10^4$, the drag force ($\overrightarrow{F}_{d}$) exerted by the air on the projectile is proportional to the square of the velocity in the form $\overrightarrow{F}_{d} \propto -\lvert \overrightarrow{v} \rvert\overrightarrow{v}$. Hence, the dynamic model is given by
\begin{equation}
	m\frac{d}{dt}\overrightarrow{v}=-\frac{C_d}{2} \rho A \lvert \overrightarrow{v} \rvert\overrightarrow{v} + m\overrightarrow{g} \label{Eq:Principal}
\end{equation} where $\overrightarrow{v}=v_x(t)\hat{\imath}+ v_y(t)\hat{\jmath}$ denotes the velocity of the spherical projectile, $m$ and $A=(\pi/4) D^2$ are the mass and cross-sectional area of the sphere, respectively, $C_d$ account for drag coefficient, and $\overrightarrow{g}=-g\hat{\jmath}$ is the gravity acceleration with $|\overrightarrow{g}|=g=9.81$ $m/s^2$. Thus, the equations of motion for velocity components of a rubber bullet are given by
\begin{subequations}
	\label{Eq:PT_TLs}
	\begin{alignat}{2} \label{Eq:PT_TLs_b}
		\frac{d v_x}{dt}=&-\alpha \sqrt{v_x^2 + v_y^2}\ v_x , \\
		\frac{d v_y}{dt}=&-\alpha \sqrt{v_x^2 + v_y^2}\ v_y - g %
	\end{alignat}
\end{subequations}here $\alpha=(C_d/2m)\rho A$, the drag coefficient is $C_d=0.5$\cite{Landau} (for bodies of spherical shape), and the initial conditions for velocity components are $v_{x0}=v_0\cos\theta_0$ and $v_{y0}=v_0\sin\theta_0$.

The terminal velocity is calculated by eliminating the temporal derivatives in Eq. \eqref{Eq:PT_TLs}, i.e., $d v_x/dt=0$ and $d v_y/dt=0$   which leads to
\begin{equation}
	 V=\sqrt{\frac{2mg}{C_d\rho A}}.
	 \label{Eq:Terminal_Velocity}
\end{equation}
Replacing density value $\rho = 1.22$ $kg/m^3$ and viscosity $\mu=1.50\times 10^{-5}$ $m^2/s$ of air, the mass $m=0.64$ $g$ and diameter $D=8\times10^{-3}$ $m$ of a shot which is considered as a sphere, it is found that the terminal velocity is  $V=20.24$ $m/s$. The density and the viscosity of the air were considered under conditions of 20º C and $1$ $atm$ of pressure. Then, the corresponding Reynolds number \eqref{Eq:Reynolds_number} is equal to $Re=1.3\times10^{4}$, which confirms the assumption of the model \eqref{Eq:Principal}.

Importantly, there is a set of forces that were not taken into account within the framework of this model; due to their influence can be negligible for a first approach. Some forces are \cite{Pantaleone}: i) the buoyancy force, ii) the Magnus force, iii) the possible presence of wind, iv) the history force and related force with the added mass, and v) the centrifugal and Coriolis forces that take into account the non-inertia of the Earth's frame of reference. Here the model ends. Analysis and considerations to address in the classroom are reviewed below. 

Equations \eqref{Eq:PT_TLs}(a)-(b) allow finding, by numerical integration regarding time, the velocities $v_x$ and $v_y$, and further the position of the projectile on the $x-axis$ and the $y-axis$. Students can use this model to analyze different cases at straight, and low angle shots, placing forward approaches to evaluate relevant physical parameters in a rubber bullet impact generating body injuries. In this case, the analyses are focused on ocular trauma in conformity with the arguments of section II.

In a first analysis, calculations used an initial angle of injection of $\theta_0=0^{\circ}$, and initial velocities of $v_0= 320$ $m/s$ and $v_0= 380$ $m/s$ corresponding to the supplier and the police reported data, respectively. Moreover, it is assumed that the projectile is fired horizontally at an initial height of $1.7$ $m$ measured from the floor, disregarding the dispersion due to the transmitted momentum of the other pellets. Other studies\cite{Chandranath,Nag} detail essential considerations to understand various dispersion models of cartridge bullets fired to multiple distances. It should also be noted is regarding the time of phenomena description; all analyses concern the time taken from the exit of the shotgun riot to the target, leaving aside the physical aspects when the bullet is inside the barrel.

\begin{figure}[t!]
 \centering
\includegraphics[width=1\textwidth]{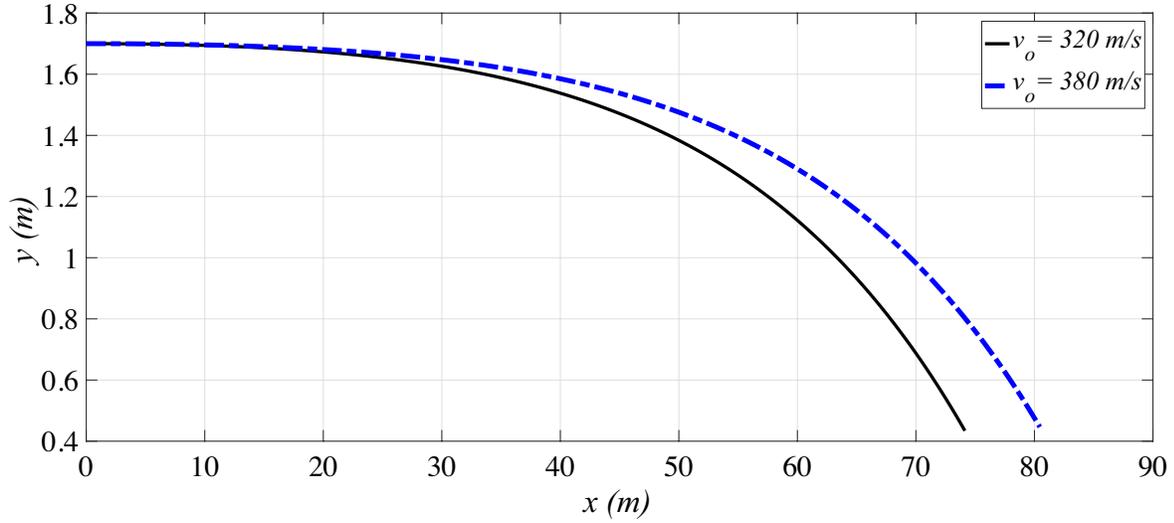}

\caption{Projectile trajectory graph, $y=f(x)$. There were used initials velocities of $v_0=320$ m/s and $v_0=380$ $m/s$ with a firing angle of $\theta=0^{\circ}$.}
\label{Fig:Trajecetory}
\end{figure}

Figure \eqref{Fig:Trajecetory} shows the trajectory of the rubber bullet; the curve of the height versus the $x$-position shows how the bullet begins to descend, losing altitude. 
As can be seen, for the initial speed of $320$ $m/s$ and $380$ $m/s$, the projectile has descended only $16$ $cm$ and $11$ $cm$, respectively, when it travels $40$ $m$. It is possible to define correlations between reported data in the literature for eye damage in terms of velocity, energy, and energy per unit area of impact at different shooting distances. This information would be valuable to establish riot shotgun use protocols, public prevention, medical records, forensic examination cases, among others. Therefore, if a standing person suffers severe ocular trauma, in the present case, it should have been at some point in the projectile trajectory. At this point, the model defines a particular value of the physical parameters mentioned previously.

\begin{figure}[t!]
 \centering
\includegraphics[width=1\textwidth]{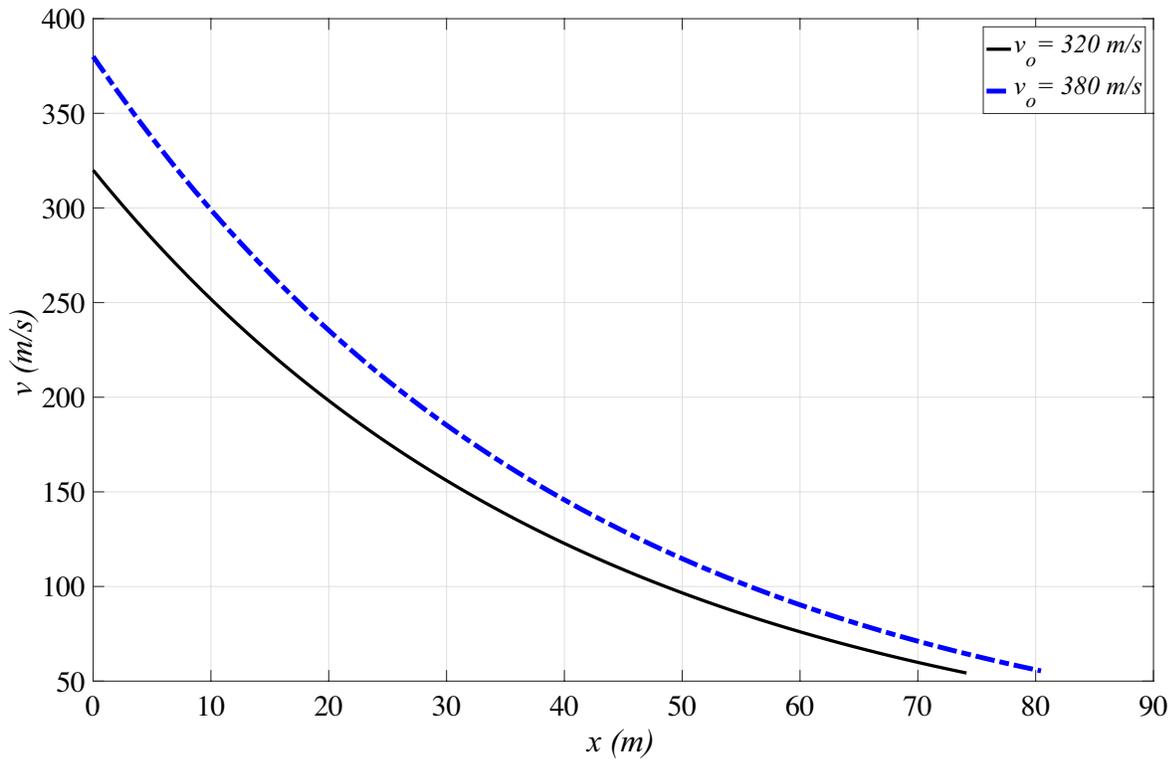}
\caption{Velocity $v=\sqrt{v_x^2+v_y^2}$ as a function of position $x$ for a rubber bullet fired at $\theta=0^{\circ}$ with initial velocities of $v_0=320$ $m/s$ and $v_0=380$ $m/s$.}
\label{Fig:Velocidad}
\end{figure}

Figure \eqref{Fig:Velocidad} shows how the speed varies depending on the distance, given the influence of the medium. Similar behaviors will have the kinetic energy depending on the projectile mass and the square of the velocity, $E_c=m v^2/2$. A rubber bullet of mass $0.64$ $g$ fired at an initial speed of $320$ $m/s$ will have a kinetic energy of $32.8$ $J$, in the case of the same projectile with an initial speed of $380$ $m/s$ the kinetic energy will be $46.2$ $J$. In both cases, the single pellet energy is below the energy limit of $122$ $J$. 
However, since the cartridge contains 12 bullets with similar energy, assuming the same structural and physical conditions, the riot gunshot weapons exceed the limits being lethal at short distances.

Reviewing the firing length suggested, which is $30$ $m$, the bullets will have decreased their speed approximately to $154$ $m/s$ and $180$ $m/s$ when they have been fired at initial speeds of $320$ $m/s$ and $380$ $m/s$, respectively. The projectiles have reduced their speed by more than half for both cases; they will have a kinetic energy of $7.5$ $J$, and $10.4$ $J$, respectively. Considering the 12 bullets of a cartridge, the kinetic energy will be $90$ $J$, and $124.8$ $J$, magnitude that although their current assessment, is very close to the internationally recommended value of $122$ $J$. This fact would suggest that $30$ $m$ is a safe distance that significantly reduces the probability of an impact causing personal injuries. However, before taking this statement for granted, some considerations have to be made. First, the value of $122$ $J$ is a reference, not an exact limit that discriminates between severe or mild damage. Besides, essential characteristics as the caliber of the ammunition, mass, bullet dimensions, or materials of the pellets used to define this value are different from those studied in this case. For these reasons, defining a degree of injury only by the kinetic energy of the shot would be unsuitable. 

\begin{figure}[b!]
\centering
\includegraphics[width=0.95\textwidth]{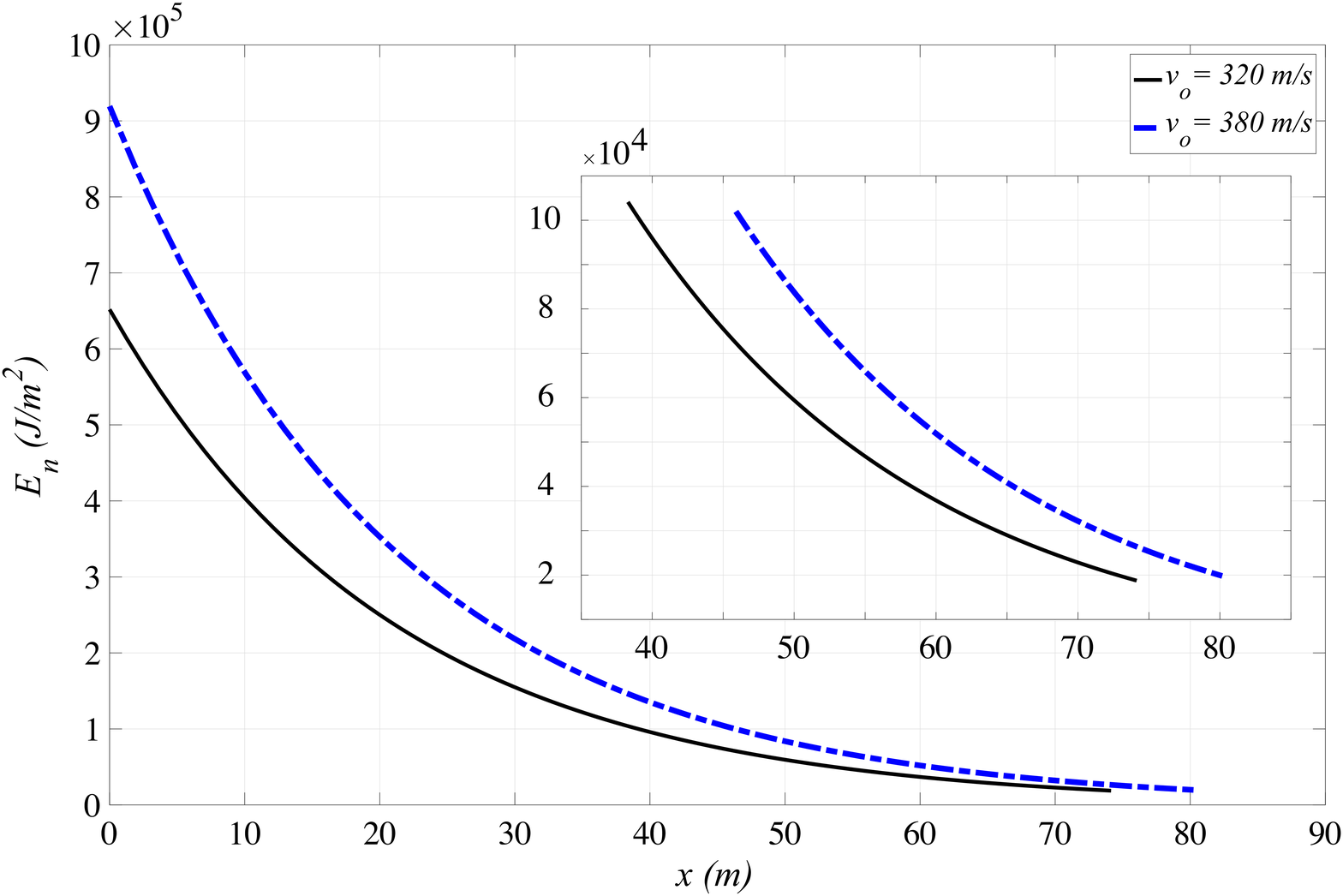}
\caption{Normalized energy $E_n$ as a function of position $x$ for a rubber bullet fired at $\theta=0^{\circ}$ with initial velocities of $v_0=320$ $m/s$ and $v_0=380$ $m/s$.}
\label{Fig:Energia_x}
\end{figure}

In contrast, if the kinetic energy is expressed in terms of the cross-section of the projectile, the students will be capable of performing correlations between possible human body injuries and normalized energy impacts. This comparison is significant since a projectile with kinetic energy acting over a smaller impact section will have higher bones and tissues drilling capacity. In the present case, for rubber bullets, the projected area will be a circle of radius 8 $mm$. Thus, the energy is normalized by their cross-sectional area and then compared with reported values in literature obtained from experimental investigations for ocular trauma produced by similar kinetic projectiles. Figure \eqref{Fig:Energia_x} shows the behavior of kinetic energy per unit area through distance, and the box shows a zoom of the earliest values. Realize that to $30$ $m$ impacted person on the face will absorb the kinetic energy of approximately $165000$ $J/m^2$ and $210000$ $J/m^2$ when projectile has been fired at an initial speed of $320$ $m/s$ and $380$ $m/s$, respectively. 

On the one hand, information  provided in 2012 states that to $30$ $m$ there exist globe rupture probability. According to the literature, a 50\% risk of globe rupture occurs just over $36000$ $J/m^2$. Curves show projectile reaching this value at a distance near $60.4$ $m$ and $67.6$ $m$ away when projectiles are fired at $320$ $m/s$ and $380$ $m/s$, respectively. As previously discussed, the energies per unit area of other eye injuries are $17979$ $J/m^2$ for 50\% risk of retinal damage, over $17300$ $J/m^2$ for 50\% risk of lens damage, and over $11700$ $J/m^2$ for 50\% risk of hyphema. However, these energy values are reached at over 75 m and 85 m, respectively; almost 2.5 and 3 times over recommended fired distance to avoid injuries. This evidence is necessary to discuss recommendations for use.

On the other hand, it is also possible to supplement the analysis simulating what happens with the impact energy, if the projectile bounced off the ground before reaching its target. Based on the above model, it has been estimated the trajectory of a projectile bouncing in an inelastic collision after being fired with an initial angle of $\theta=2^{\circ}$ below the horizontal to $320$ $m/s$. The coefficient of restitution of rubber is around 0.8. It is recognized that this parameter depends on properties such as stiffness, toughness, strength, and hardness of the two bodies involved in the collision. So it is worth mentioning that in the introductory physics courses, the microscopic composition of ground and projectile to describe the kinetic energy after a bounce is neglected. However, further courses, as the science of materials, will allow the student to address a rigorous description of it. Thus for the estimate, the model uses the values of 0.4 to 0.9 for this parameter. These cases can describe a deterrence situation where rubber bullets go straight to the ground.

\begin{figure}[t!]
 \centering
\includegraphics[width=0.98\textwidth]{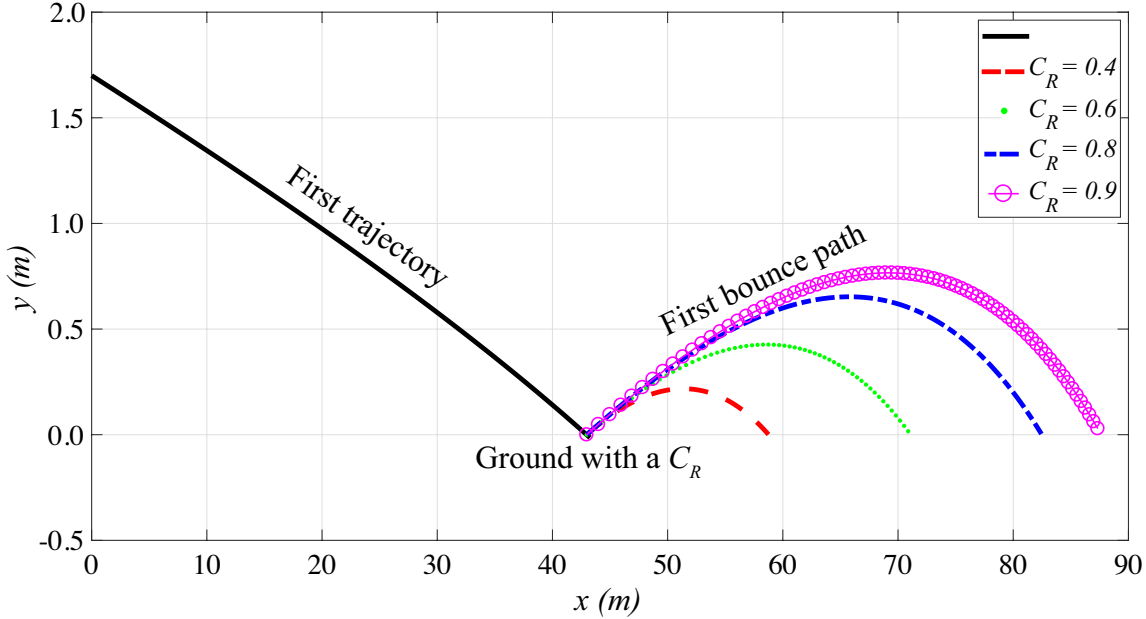}
\caption{The trajectory of a projectile fired below the horizontal line producing one bounce. The initial velocity of the rubber bullet is $v=320 $ $m/s$, and the firing angle is $\theta=-2^{\circ}$.}
\label{Fig:rebote}
\end{figure}

\begin{figure}[h!]
 \centering
\includegraphics[width=0.98\textwidth]{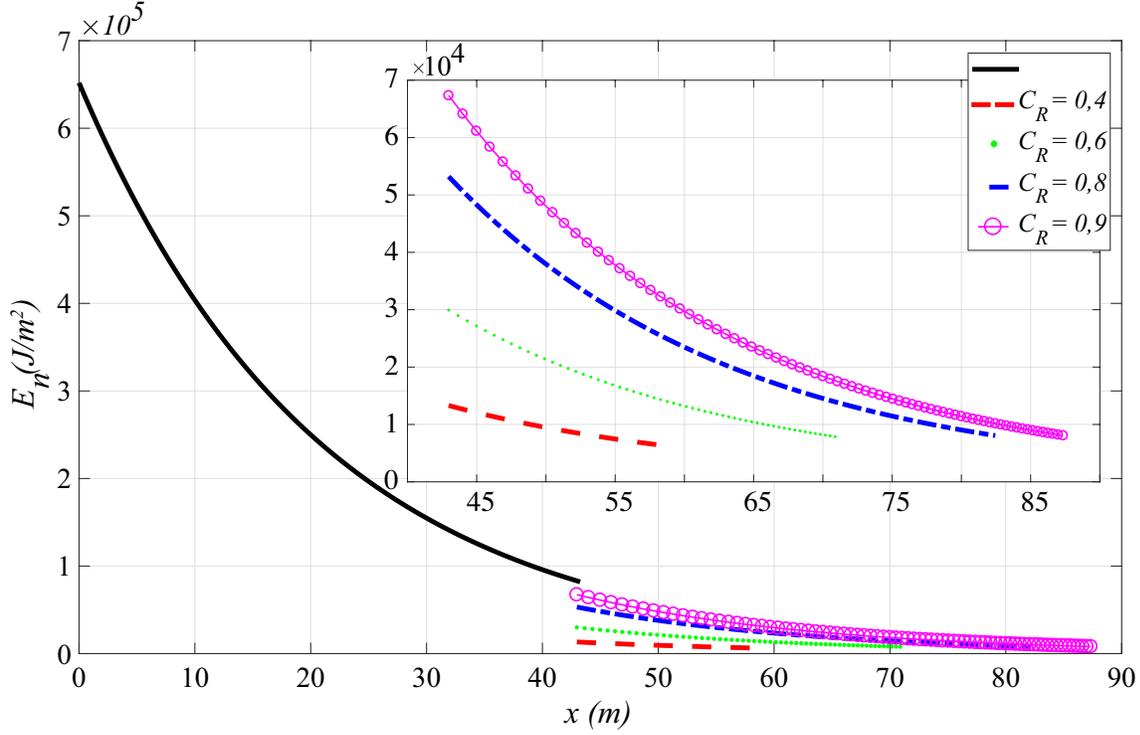}
\caption{{The normalized energy of a projectile fired below the horizontal line producing one bounce. The initial velocity of the rubber bullet is $v=320 $ $m/s$, and the firing angle is $\theta=-2^{\circ}$.}}
\label{Fig:Normalized energy}
\end{figure}

To analyze the bounce trajectory; initially, the model describes the motion until the projectile impacts with the ground; then, the velocities and energies of the bullet before and after the collision are calculated using the restitution coefficient. Finally, the bounce angle, getting from the inverse tangent of the velocity components relationship, and the velocity after ground impact will be the new initial conditions for applying the model one more time. Figure \eqref{Fig:rebote} shows that the maximum extent in horizontal and vertical direction decrease with smaller coefficients of restitution. Likewise, the normalized energy has a similar curve tendency but different values in each bounce case, as is shown in figure \eqref{Fig:Normalized energy}.


To realize that from the energy curve, even ricochets have enough energy to cause eyeball injuries, even ocular rupture. However, these values must be correlated with the trajectory to define if these projectiles hit the face or other body parts. In the latter case, the normalized energies must be those that cause specific damage in bone, skin, limbs, among others. 

Table \ref{tabla_1} shows normalized energy and vertical height reached by a projectile to distances between $45$ $m$ and $80$ $m$ for restitution coefficients 0.4, 0.6, 0.8. and 0.9.
After the bounce, the projectile reaches a maximum height of $76.5$ $cm$ for a $C_R$=0.9 having an energy of $18398.7$ $J/m^2$. If people are $70$ $m$ away from the shooter, under these conditions, the projectile would not strike the victim's face but could hurt a child. 

\begin{table}[b!]
\centering
\caption{Normalized energy and maximum height after a rebound; $v=320$ $m/s$ and $\theta=-2^{\circ}$.}
\begin{center}
\begin{tabular}{ccccc}
\hline
$C_R$ & 0.4 & 0.6 & 0.8 & 0.9\\
$x(m)$
 & \multicolumn{1}{p{3.5cm}}%
{\centering $E_n(J/m^2)$ \hspace{0.2cm}{$y(cm)$}}
 & \multicolumn{1}{p{3.5cm}}
{\centering $E_n(J/m^2)$ \hspace{0.2cm}{$y(cm)$}}
 & \multicolumn{1}{p{3.5cm}}%
{\centering $E_n(J/m^2)$ \hspace{0.2cm}{$y(cm)$}}
 & \multicolumn{1}{p{3.5cm}}%
{\centering $E_n(J/m^2)$ \hspace{0.2cm}{$y(cm)$}}
 \tabularnewline 
\hline
45 & \hspace{-0.2cm}12200.0\hspace{1.0cm}7.9 & \hspace{-0.0cm}27174.1\hspace{0.9cm}9.3 & \hspace{-0.0cm}48827.1\hspace{0.9cm}8.5 & \hspace{-0.0cm}61154.9\hspace{0.9cm}9.6\\
50 & 9612.0\hspace{0.8cm}20.8\hspace{-0.1cm}& 21443.6\hspace{0.8cm}28.1 & 38497.4\hspace{0.8cm}29.8 & 48958.6\hspace{0.8cm}30.0\\
55 & 7527.7\hspace{0.8cm}17.3\hspace{-0.1cm}& 16927.3\hspace{0.8cm}39.7 & 30109.6\hspace{0.8cm}47.5 & 38592.1\hspace{0.8cm}48.6\\
60 & -\hspace{1.6cm}-\hspace{-0.4cm} & 13407.3\hspace{0.8cm}42.4 & 23492.8\hspace{0.8cm}59.9 & 30177.8\hspace{0.8cm}64.0\\
65 & -\hspace{1.6cm}-\hspace{-0.4cm} & 10474.6\hspace{0.8cm}33.1 & 18843.6\hspace{0.8cm}65.0 & 23542.6\hspace{0.8cm}73.8\\
70 & -\hspace{1.6cm}-\hspace{-0.4cm} & \hspace{0.1cm} 8282.6\hspace{1.0cm}9.1&
14751.9\hspace{0.8cm}62.4 & 18398.7\hspace{0.8cm}76.5\\
75 & -\hspace{1.6cm}-\hspace{-0.4cm} & -\hspace{1.6cm}-\hspace{-0.4cm} & 11632.0\hspace{0.8cm}49.1 & 14781.0\hspace{0.8cm}71.4\\
80 & -\hspace{1.6cm}-\hspace{-0.4cm} & -\hspace{1.6cm}-\hspace{-0.4cm} & \hspace{0.3cm}9253.1\hspace{0.8cm}23.4 \hspace{-0.1cm}& 11655.0\hspace{0.8cm}55.6\\
\hline
\end{tabular}
\end{center}
\label{tabla_1}
\end{table}

It is noteworthy to mention that there are energy parameters that should impact the eye produce lens and retina injuries, but most cases without globe rupture. This fact means that the projectile behaves as a blunt object causing a closed globe injury \cite{Kuhn}. After the bounce, there is only likely to produce global rupture for a $C_R =0.8$ and $C_R=0.9$ at a horizontal distance of $50$ $m$, and $55$ $m$, respectively, when values of normalized energy take values higher than $36000$ $J/m^2$. For other combinations of position and $C_R$, will result in blunt blows without globe rupture. The bounce height ranges from $8$ $cm$ and $80$ $cm$ depending on the coefficient of restitution and properties of the ground and projectile. 

\begin{table}[t!]
\centering
\caption{Normalized energy and maximum height after a rebound; $v=380$ $m/s$ and $\theta=-2^{\circ}$.}
\begin{center}
\begin{tabular}{ccccc}
\hline
$C_R$ & 0.4 & 0.6 & 0.8 & 0.9\\
$x (m)$
 & \multicolumn{1}{p{3.5cm}}%
{\centering $E_n(J/m^2)$ \hspace{0.2cm}{$y(cm)$}}
 & \multicolumn{1}{p{3.5cm}}
{\centering $E_n(J/m^2)$ \hspace{0.2cm}{$y(cm)$}}
 & \multicolumn{1}{p{3.5cm}}%
{\centering $E_n(J/m^2)$ \hspace{0.2cm}{$y(cm)$}}
 & \multicolumn{1}{p{3.5cm}}%
{\centering $E_n(J/m^2)$ \hspace{0.2cm}{$y(cm)$}}
 \tabularnewline 
\hline
45 & 17196.0\hspace{1cm}2.3 & 38216.8\hspace{1cm}3.5 & 67109.9\hspace{1cm}4.7 & 84416.8\hspace{0.8cm} 5.2\\

50 & 13582.3\hspace{0.8cm}18.9& 29912.7\hspace{0.8cm}23.6 & 53195.3\hspace{0.8cm}25.0 & 68453.0\hspace{0.8cm}23.9\\

55 & 10579.7\hspace{0.8cm}24.3& 24049.0\hspace{0.8cm}37.0 & 41538.7\hspace{0.8cm}43.5 & 54143.7\hspace{0.8cm}42.7\\

60 &8486.2\hspace{0.8cm}15.7\hspace{-0.2cm} & 18741.8\hspace{0.8cm}45.7 & 33332.4\hspace{0.8cm}56.7 & 42194.3\hspace{0.8cm}59.6\\

65 & -\hspace{1.6cm}-\hspace{-0.4cm} & 14678.8\hspace{0.8cm}35.9 & 26499.3\hspace{0.8cm}66.2 & 32653.1\hspace{0.8cm}72.9\\

70 & -\hspace{1.6cm}-\hspace{-0.4cm} & 11816.9\hspace{0.8cm}35.9 & 20984.3\hspace{0.8cm}70.3 & 26019.2\hspace{0.8cm}80.0\\

75 & -\hspace{1.6cm}-\hspace{-0.4cm} & 9212.9\hspace{0.7cm} 11.1\hspace{-0.1cm} & 16221.2\hspace{0.8cm}66.6 & 20647.9\hspace{0.8cm}81.8\\

80 & -\hspace{1.6cm}-\hspace{-0.4cm} & -\hspace{1.6cm}-\hspace{-0.4cm} & 12923.9\hspace{0.8cm}54.0 & 16384.8\hspace{0.8cm}76.4\\
\hline
\end{tabular}
\end{center}
\label{tabla_2}
\end{table}

Similarly, table \ref{tabla_2} shows the normalized energy and maximum height reached by a projectile fired with $\theta=-2^{\circ}$ and an initial velocity of $380$ $m/s$. In this case, when the restitution coefficient is 0.9, the projectile reaches a maximum height of $81.8$ $cm$ at $75$ $m$ from the shot. As expected, when the speed increases, the number of cases in which an open globe eye injury occurs is higher. Even more, for larger angles below the horizontal; the bounce distance will be closer to the firing officer; therefore, both bounce heights, and impact energy will be more significant, thus increasing the likelihood of an open globe injury. These are excellent exercises for the students to work out these trends for themselves. 

At this point, the teacher can ask students some further variants for the problem, such as:
\begin{itemize}
 \item Investigate biomechanical properties of the skin or other tissue to determine possible body injuries in terms of normalized energy values. Depending on the course or education, studies and analyses can be performed experimentally or by using experimental data from the literature.
 \item To use the analysis model for other projectiles or materials, implying that after the ricochet, higher lengths are reached, and the probability of eye damage persists.
 \item Develop experimental activities to simulate the case of projectiles, with large balls thrown at low speeds to avoid damage. Thus, one can experimentally test the relationship between firing angles, horizontal and vertical range achieved by the projectile. 
 \item To implement active learning with modeling instruction, favorable attitude changes are obtained in the introductory courses at university level\cite{Brewe,LaGarza}.
\end{itemize}

The experiences described above presented a detailed discussion of the pellet motion with quadratic resistance to air; to accomplish this purpose, the student must develop the necessary skills to operate any programming language that supports mathematical calculations to solve equations \eqref{Eq:PT_TLs}. The teacher must encourage their students to do this on a cross-cutting basis. On the other hand, the teacher must instruct how to work out the movement problem analytically; thus, the student can evaluate the scope of the proposed model and their considerations at the time of its description. Accordingly, by addressing the launch of the projectile considering small angles $\theta$, it is correct to assume that the horizontal velocity is, on average, much higher than the vertical velocity, that is, $v_x(t) \gg v_y(t)$. Therefore in this approximation, the equations of motion \eqref{Eq:PT_TLs} are given by
\begin{subequations}
	\label{Eq:PT_TLs_small}
	\begin{alignat}{2} \label{Eq:PT_TLs_smallb}
		\frac{d v_x}{dt}=&-\alpha v_x^2.\\
		\frac{d v_y}{dt}=&-\alpha v_x v_y - g. %
	\end{alignat}
\end{subequations}

Thus, considering the initial conditions $x(0)=x_0$, $y(0)=y_0$, and by integrating directly \eqref{Eq:PT_TLs_small}, the closed-form solutions for the velocity can be written as:
\begin{subequations}
	\label{Eq:PT_TLs_solution_a}
	\begin{alignat}{2} \label{Eq:PT_TLs_solution_a}
		v_x(t)&= \frac{1}{(\alpha t +1/v_{x0})},\\
		v_y(t)&=\frac{\tan(\theta_0)-((\alpha/2) t + 1/v_{x0})gt}{\alpha t + 1/v_{x0}},
	\end{alignat}
\end{subequations}
and for the position as:
\begin{subequations}
	\label{Eq:PT_TLs_solution_b}
	\begin{alignat}{2} \label{Eq:PT_TLs_solution_b}
        x(t)&= x_0+ \frac{\ln(\alpha v_{x0} t+ 1)}{\alpha},\\
        y(t)&=y_0 -\left(\frac{2}{\alpha v_{x0}}+t\right)\frac{gt}{4}  +\left( \frac{g}{2\alpha v_{x0}}+ v_{y0}\right)\frac{\ln(\alpha v_{x0} t+1)}{\alpha v_{x0}}.
	\end{alignat}
\end{subequations}

These analytical solutions for low angles will allow students to analyze multiple projectile launches to evaluate potential body damage when an impact occurs.

\begin{figure}[t!]
 \centering
\includegraphics[width=1\textwidth]{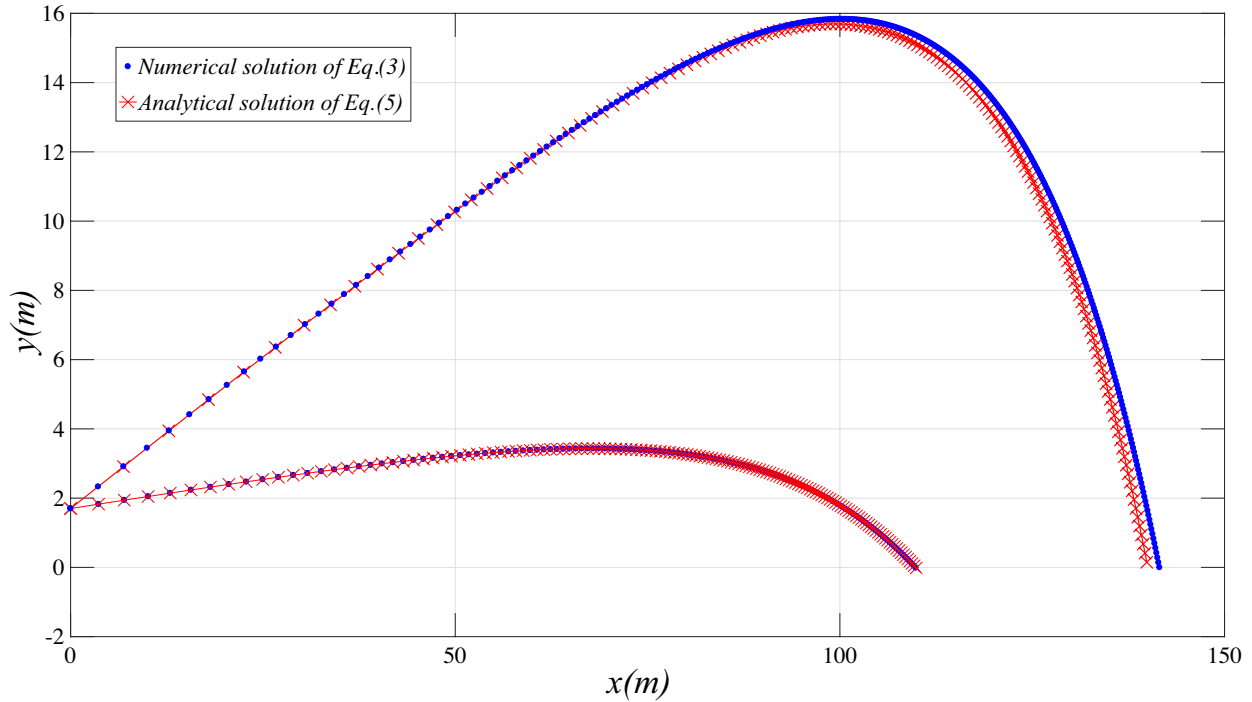}
\caption{Comparison between two trajectories obtained from the numerical solution of Equation \eqref{Eq:PT_TLs} and the analytical solution of Equation \eqref{Eq:PT_TLs_small}.}
\label{Fig:comparison_xy}
\end{figure}

Figure \ref{Fig:comparison_xy} shows the similarities between analytical and numerical solutions under the approximation of small angles. Thus, both solutions allow students to describe the movement of the projectile correctly while analyzing the physical parameters associated with eye injuries.

\section{Conclusions}
An analysis of a real case contextualized to promote the use of evidence in physics class was presented. The obtained results using basic concepts of introductory physics represent reliable support for the analysis of eye damage from the correlation between the energy per unit area of impact on the surrounding region about the orbital cavity housing the eyeball.

For this purpose, it was implemented a first scientific exploration of the cinematic of a spherical projectile using coupled ordinary differential equations ($ODEs$) built based on the value of the Reynolds number, $Re\gg1$ . The $ODEs$ were integrated numerically to find the speed and position of the projectile, allowing to compare the energy per unit of area with reported values for ocular injuries. Finally, since there is a wide range of possible shots, it is proposed to make an approximation for the description of the movement for small angles giving students an essential tool to discern over other projectile movements.

Projectile kinetic impact cases with official reported data about the type of weapons used to fire rubber bullets of $8$ $mm$ in diameter and $0.64$ $g$ were analyzed. The study takes into account air friction, air density, and loss of height due to gravity. The impact energy per unit area is explained in the cases of straight and angle-shot, 0º, and -2º, respectively;  the latter case contemplates a rebound with different restitution indexes. These analyses conclude that the shotguns should not be used to directly shoot at distances of less than $70$ $m$ at risk of causing open or closed ocular globe injuries when a direct or indirect impact by rebound occurs.

The study also shows that is necessary to use normalized energy data and correlate it  with the biophysical characteristics of the physiology of each part of the human, front, and rear body, considering the location of the possible impact, to determine if a weapon to determine whether a weapon may be appropriate or not for use in protests. Studies must review not only the evidence of direct shot but for angled paths with positive and negative inclination in terms of the horizontal and vertical distances, as well as the coefficient of restitution at projectile rebound.

It is encouraged for tertiary education to address the same problem with other complexity and approaches. For example, considering that a single cartridge contains 12 projectiles that interact with each other at the time of firing, it is suggested to approach the analysis of the proposed problem in more advanced courses of statistical mechanics for thorough detail and complexity. Also, the analysis could deepen further in the course of physics or biophysics and health sciences.

This contribution is intended for students to discuss real cases in the classroom by applying the necessary basic physics. Without a doubt, suggesting evidence-based recommendations allows you to evaluate the use of scientific knowledge as an essential part of decision making.



\end{document}